Oxygen atoms and molecules at $La_{1-x}Sr_xMnO_3$ surfaces

Walter A. Harrison
Applied Physics Department
Stanford University
Stanford, CA, 94305-4045

Abstract

A localized description, rather than energy bands, is appropriate for the manganite substrate. Empty substrate levels lower in energy than occupied oxygen levels indicate need for further terms beyond the Local Density Approximation. So also does van-der-Waals interaction between the two. Methods to include both are suggested by related, exactly soluble, two-electron problems. The descriptions of the electronic structure of the molecule and a $La_{1-x}Sr_xMnO_3$ (LSM) substrate are greatly simplified to allow incorporation of these effects and to treat a range of problems involving the interactions between oxygen atoms, or oxygen molecules, and such a substrate. These include elastic impacts, impacts with electronic transitions, and impacts with phonon excitation. They provide for capture of the atoms or molecules by the surface, leaving the neutral molecule strongly bound over a $Mn^{4+}$ site. It is found that oxygen vacancies in LSM diffuse as a neutral species, and can appear at the surface. Bound molecules tend to avoid sites next to vacancies but, if there, should drop one atom into the vacancy leaving the remaining triplet oxygen atom bound to the resulting ideal surface, with no need for spin flips nor successive ionization steps.

1. Introduction

The absorption and uptake of oxygen on manganites is central to the behavior of oxide-based fuel cells. There have been a number of calculations using Density Functional Theory (DFT) for treating oxygen and other molecules near surfaces[1-3], mostly for metals[4-6]. They seemed not to address the energy-loss mechanisms which we wished to understand. We therefore sought to explore the interaction of oxygen with a surface of $La_{1-x}Sr_xMnO_3$ (LSM) in terms of the simplified descriptions of electronic structure such as described in Ref. 7, but using the localized description for the manganites which had proven successful in Refs. 8 and 9. We quickly learned that the DFT had additional difficulties with this particular system, which might be less obvious when using a full computer code to treat the electronic structure.

The first difficulty arises with an oxygen atom, for which the occupied $p$ state has an energy significantly higher than the lowest empty states in the LSM. Thus in DFT electronic charge would be transferred from the oxygen atom to the substrate, even if it was far away, until that level dropped to the energy of the empty LSM levels. This might not be serious when the atom is close to the surface, but it leads to quite incorrect results at large distances and it will be important to have the relative energy between these two regimes. The second difficulty was with the $O_2$ molecule for which we found in one-electron theory a repulsion between the molecule and the substrate at all distances. A net attraction arose, however, if we included the van-der-Waals interaction, which arises only beyond one-electron theory. We sought to resolve both difficulties by treating two-electron problems which have the same difficulties, but which could be solved exactly. With this as a guide we were able to proceed to a detailed, though approximate, description of this system. This also led to a form for the van-der-Waals interaction dependent only upon the geometry of the molecule, independent of the parameters we use.

We shall begin with a simple description of the oxygen molecule, following Ref. 7 but adjusting the parameters which enter to give the observed internuclear distance and binding energy. We then represent the electronic structure of the substrate in terms of cluster orbitals, based on individual Mn ions and their nearest-neighbor oxygen ions, as were used earlier[8] in the treatment of the oxides of Mn and Fe and in the generalization to LSM in Ref. 9, where we calculated a wide range of properties . At that stage we shall do the relevant two-electron problem to see how to treat the coupling between the two systems, and obtain the energy as a function of separation. Finally we use this energy to understand the behavior of an incoming oxygen atom and an incoming oxygen molecule, including the possibilities of electronic transitions in the process, and the generation of phonons in the substrate. The determining factor for phonon generation turned out to be whether there was a sufficiently strong attractive interaction with the surface to cause a sharp impact.

## 2. The Oxygen Molecule

We shall need the electronic structure of the molecule, and it clarifies the nature of our approximate description of the electronic structure of the substrate. The important electronic states of the oxygen atom are the $2p$ states at[7] $\varepsilon_p = -16.77$ eV. They are coupled in the molecule by a $V_{pp\sigma}$ and a $V_{pp\pi}$ which in Ref. 7 were taken to vary with spacing $d$ as $1/d^2$ with universal coefficients obtained from semiconductor bands. Here we need them over such a large range of $d$ that we should fit the $1/d^2$ to an exponential $\exp(-\mu d)$ which fits the $1/d^2$ at the O-O spacing in LMO with $\mu = 0.72/\text{Å}$. We then adjusted the coefficients to give the observed oxygen molecular spacing as $d_O = 1.22$ Å and the binding as 5.2 eV. The energy calculation included a repulsion, approximately proportional to the square of the coupling as in Ref. 7, and arising from an upward shift of the molecular levels due to the nonorthogonality of atomic orbitals on neighboring sites. In this study it will be important to explicitly include this by shifting the levels, which was not necessary in the problems treated in Ref. 7. Here the shifts cause levels to cross, allowing electronic transitions.   For $V_{pp\pi}/V_{pp\sigma}$ equal to $-1/3$ the $\sigma$ levels shift nine times as much as the $\pi$ levels. The couplings between two oxygen atoms which this leads us to are

$$V_{pp\sigma} = 3.9 \exp(-(r-d_O)\mu) \text{ eV},$$

$$V_{pp\pi} = -1.3 \exp(-(r-d_O)\mu) \text{ eV},$$

(1)

and the shifts due to nonorthogonality become

$$\delta\varepsilon_p = \lambda V_{ppm}^2/|\varepsilon_p|$$

(2)

with $\lambda = 2.15$ and with $m$ either $\sigma$ or $\pi$. There is also a shift of the minority-spin states relative to the majority-spin states in the paramagnetic molecule due to an exchange energy given by $U_x = 2.34$ eV (the energy lowering for each pair of $p$ states with parallel spin on one oxygen atom, obtained from the NIST tables of atomic spectra). This shifts minority-spin level in the atom by $2U_x$ but the molecular levels by only $U_x/2$, as shown for the ground states of each, to the middle and right in Fig. 1.

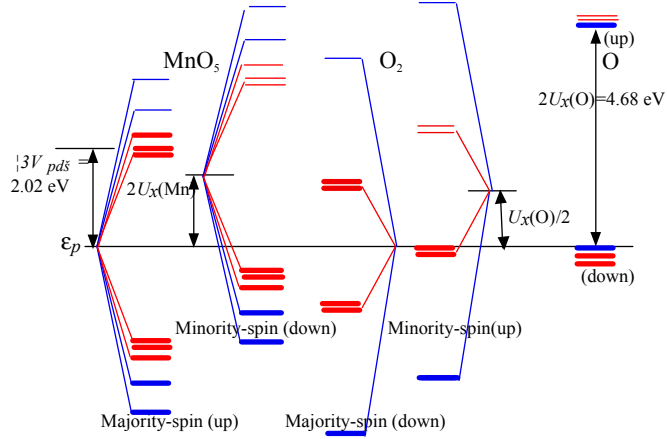

Fig 1. Energy levels, all measured from the oxygen majority-spin $p$-state energy, for the MnO$_5$ surface cluster in SrMnO$_3$ on the left, for an oxygen molecule in the middle, and for the oxygen atom on the right, showing the majority-spin and minority-spin levels for each. Each line represents one level; closely-spaced lines are degenerate levels. $e_g$ levels for MnO$_5$ and $\sigma$-levels for O are shown in blue, $t_g$ levels for MnO$_5$ and $\pi$ levels for O are shown in red. Those occupied in the ground state for the neutral oxygen atom, molecule, and for the Mn$^{4+}$ clusters in SrMnO$_3$ are drawn heavy. In LaMnO$_3$ the lower of the two upper majority-spin $e_g$ states is also occupied.

## 3. The Manganite Substrate

In a study of manganese and iron oxides[8] we found that the important states, the counterpart of $p$ states for the oxygen molecule, could be described as cluster orbitals, based upon the $d$ states on the Mn and the $p$ states on the six neighboring oxygen ions, MnO$_6$ clusters. In an energy-band description this corresponded to use of a *special points* method[7] of sampling the band, and in the more appropriate localized description it corresponded to localized cluster states. The same description applies to manganites in the perovskite structure[9]. The coupling between Mn $d$ states and oxygen $p$ states was taken as[7-9]

$$V_{pd\sigma} = -(3\sqrt{15}/2\pi)\hbar^2(r_d^3 r_p)^{1/2}/md^4, \qquad (3)$$

and $V_{pd\pi} = -V_{pd\sigma}/\sqrt{3}$. With parameters from Ref. 7 this leads to $V_{pd\sigma} = -2.02$ eV for SrMnO$_3$ with $d = 1.90$ Å. It is slightly smaller in LaMnO$_3$ due to a larger spacing but we neglect the difference. The variation as $1/d^4$ is sufficiently rapid that we used it throughout, with no replacement by an exponential.

In the manganites the minority $d$ states are shifted from the majority states at $\varepsilon_d = -17.22$ eV (close enough to $\varepsilon_p$ of $-16.77$ eV that we take it equal to $\varepsilon_p$) due an exchange energy of[7] $U_x = 0.78$ eV. In the octahedral clusters of the bulk manganites the cluster levels of each spin are split into three $t_g$ levels at $\pm 2V_{pd\pi}$ and two $e_g$ levels at $\pm\sqrt{3}\,V_{pd\sigma}$. Surface Mn ions have fewer neighbors and the cluster orbitals are recalculated. In SrMnO$_3$ all of the lower levels (with the − sign) are occupied; the upper minority-spin levels are empty and both majority-spin upper $e_g$ level are empty; all of the majority-spin upper $t_g$ levels are occupied. LaMnO$_3$ is the same except that one of the upper majority-spin $e_g$ levels is occupied. The resulting levels for the surface clusters of the MnO$_2$ surface of the substrate were shown, along with the oxygen levels, in Fig. 1. We have not included the shifts due to nonorthogonality in Fig. 1. A substrate of La$_{1-x}$Sr$_x$MnO$_3$ has the same cluster orbitals as for SrMnO$_3$ except that a fraction $1-x$ of them have cluster orbitals based upon a Mn$^{3+}$ ion, rather than a Mn$^{4+}$ ion, with one majority-spin $e_g$ orbital occupied. [This, again, is a localized description. Density-functional theory would place them in shared bands with quite different properties.]

When a molecule approaches the substrate we add the coupling between the levels in Fig. 1, using Eq. (3), and we shall also need the shifts due to nonorthogonality. We take them to be the same form as in Eq. (2), but with $V_{pdm}$, giving a repulsion proportional to $1/d^8$ and we need to readjust $\lambda$. In studying the elastic constants we were successful if we took the repulsion between O ions and between Sr and O ions also to vary as $1/d^8$ with the same coefficient, making these repulsions smaller by a factor 16 because their bulk spacings are greater by a factor of $\sqrt{2}$. Thus we wrote the energy gain per formula unit for the bulk arising from the sum of the energies of the occupied cluster orbitals, and the repulsive terms $\lambda V_{pdm}^2/|\varepsilon_p|$ for all of these occupied orbitals and the twelve O-O and twelve O-Sr repulsive interactions. Each of these repulsions turns out to be given by

$$V_0(r) = 4\lambda V_{pd\sigma}^2/|\varepsilon_p|, \tag{4}$$

with $V_{pd\sigma}(r)$ for Mn-O repulsions evaluated at their spacing $r = d$, and for O-O and O-Sr repulsions evaluated at the $r = \sqrt{2}d$ spacing. Then the minimum occurred at the equilibrium spacing for SrMnO$_3$ if we chose $\lambda = \sqrt{3}|\varepsilon_p/V_{pd\sigma}|/10 = 1.44$. For the oxygen approaching a substrate, Mn-O repulsions were incorporated as shifts in the occupied orbitals, but the other repulsions used Eq. (4) directly.

The energy for an incoming oxygen will be lowest if we match the minority spin on the oxygen with the majority spin on the cluster because it couples occupied and empty states which are closer in energy. We proceeded thus with up spin being majority on the cluster and minority on the oxygen, and down spin being reversed, both for the case of molecular oxygen and for atomic oxygen, as also indicated in Fig. 1.

## 4. The Two-Electron Model

We can see the difficulty in treating the coupling between oxygen and the substrate most clearly for the free atom of oxygen to the right in Fig. 1. The occupied up-spin level of the oxygen atom is higher in energy than a number of empty levels in the substrate. With empty levels lower than an occupied level we could imagine transferring the electron across, but in reality we can not. An electron would have to be added to the cluster at the electron-affinity level, higher by a Coulomb $U$, here reduced by the attraction to the hole that would be left on the O atom a distance $r$ away, $-e^2/r$. The same electron affinity shift by a $U$ arises for placing an electron on a neutral O atom. We did not need to worry about this for the atoms in forming the $O_2$ molecule in Section 2 because the two Coulomb terms very nearly cancel (as described in Ref. 7) at the spacing $d_O$ = 1.22 Å. Here we take that cancellation as exact so at a larger distance the effective $U$ for adding an electron to an oxygen atom is

$$U^*(O) = e^2/d_O - e^2/r \tag{5}$$

if this is positive, and we take it to be zero otherwise. We similarly take for the cluster, with Mn-O distance $d$ = 1.90Å, $U^*(Mn) = e^2/d - e^2/r$ if >0. The question is: how are these shifts to be incorporated in the calculation of levels when an oxygen approaches a substrate? The answer is simple for this *single* occupied oxygen state: we add $U^*(Mn)$ to the energy level of the cluster orbital, raising it far above the occupied level of a distant oxygen. As the oxygen approaches, the energy to occupy this cluster level would be lowered by interaction with positively charged O which would be formed if an electron were transferred. Correspondingly, the energy of that cluster orbital, entering a bond between the cluster orbital and the oxygen, drops and the bond formed shifts more and more toward the cluster, eventually lying predominantly on the cluster when the $U^*(Mn)$ reaches zero. This is a one-electron solution, but differs from DFT by the addition of the $U^*(Mn)$ to the empty level.

This is less clear when two electrons are involved as for the occupied up-spin $t_g$ and empty *molecular* oxygen $\pi$ levels in Fig. 1, which are close in energy though the occupied level is not higher than the empty level. This is a many-body problem, with energies of each electron depending upon the state of others. We can answer it for the simple two-electron problem including only these two sets of levels, which is exactly soluble. Such a two-electron problem, with two levels of the *same* energy on different sites, coupled by $V$ and with extra energy $U^*$ if both electrons are on the same site, was treated for example in Ref. 7, p. 594. It was solved for two electrons of opposite spin, based upon four two-electron states, one with both electrons in the first level, one with both on the second, and two with one electron on each. With reflection symmetry it could be reduced to a quadratic equation giving a two-electron energy,

$$E = U^*/2 - \sqrt{(U^{*2}/4 + 4V^2)}. \tag{6}$$

The system discussed here has two orbitals (e. g., $zx$ and $yz$), rather than two spin states, but the mathematics differs only in that the orbitals on the two sites have different energy. Then rather than an analytic solution, Eq. (6), we must solve the four-by-four Hamiltonian matrix numerically. We did this with the coupling from Eq. (3), $U^*(O)$, and taking the occupied $e_g$ states $\Delta$ = 1 eV higher than the empty $\pi$ states (as a test). The result was almost indistinguishable, in a plot, from twice the one-electron energies,

$$\varepsilon = (U^* - \Delta)/2 - \sqrt{(((U^* - \Delta)/2)^2 + V^2)} \tag{7}$$

(measured from the starting occupied state) over the entire range of $r$, though on close inspection it differed by some 10% at the midrange of separation $r$. In our approximate approach it is appropriate to neglect the difference and proceed in this simple one-electron way, with the principal DFT error corrected by the insertion of a $U^*(r)$.

Before proceeding with that, we should note that a closely related model suggests the form for the *van-der-Waals interaction* between the molecule and the substrate, which is not included in one-electron theory. The general theory has been given in Refs. 10 and 11, which however focused upon effects of retardation which are not important here. It may be best to make a simple direct treatment here which gives the needed result. We note that the problem of the two $\pi$ bonds in the molecules is the same as that we just did, but with $\Delta = 0$, leading to Eq. (6). We next add the effects of the image potential arising from the substrate, taking the image charge equal and opposite to the charge causing it, appropriate for a conductor or high-dielectric-constant substrate. The two basis states with one electron on each site corresponds to neutral atoms and no image charge, but the two states with both electrons on one site produces an image shifting the energy by

$$\delta U^* = \tfrac{1}{2}\,[2e^2/(2z+d_O) - e^2/2z - e^2/(2z+2d_O)] \tag{8}$$

for a molecule oriented perpendicular to the substrate at a distance $z$ from the closest atom. (The factor ½ came from the energy of a charge $e$ due to its image, $-1/2\,e^2/(2z)$.) This has exactly the effect of adding $\delta U^*$ to the $U^*$ in Eq. (6). We may then expand the square root for small $\delta U^*$ to obtain the shift in the energy of the molecule as

$$\delta E = \tfrac{1}{2}\,\delta U^*[1 - U^*/\sqrt{(U^{*2}+16V^2)}] \to \delta U^*/2. \tag{9}$$

We have taken $U^*$ as $U^*(0) = 0$ at $d_O$, to obtain the final result. This is the energy gain for two electrons involved in the bond. The more familiar form[10,11] contains the molecular polarizability, which for this simple bond is (e. g., Ref. 7, p. 147) $\alpha = e^2 d_O^2/2|V_{pp\pi}|$. Thus if we also use the limit of Eq. (8) for large $z$, which is $\delta U^* \approx -d_O^2 e^2/8z^3$, we may write the result as $\delta E = -\alpha V_{pp\pi}/8z^3$, a more familiar form but the first form ($\delta E = -d_O^2 e^2/16z^3$) is simpler and not dependent upon our particular values of parameters. The $V_{pp\pi}$ in the numerator cancels its inverse appearing in the polarizability, so that only the geometry of the molecule enters. It is an interaction varying as $1/z^3$ in contrast to van-der-Waals interactions between molecules varying as $1/r^6$ as noted in Ref. 10 and 11. (For interaction between molecules the $V_{pp\pi}$ does not cancel out.) In both references it was found that at very large distances the limited speed of light to send the image potential causes these to fall off with an additional factor of $1/z$, but we are not concerned with that regime  We note also that if we had instead placed the molecular axis parallel to the surface, the attraction would be $\delta E = \tfrac{1}{4}\,[2e^2/\sqrt{(4z^2+d_O^2)} - 2e^2/2z] \approx -d_O^2 e^2/32z^3$, half as large. The expanded form is inaccurate in either case in the region of interest and we retain Eq. (8). Further, we include it as we did for the inter-ion repulsion by adding one half (since Eqs. (8) and (9) were for a two-electron bond), $\delta U^*/4$, to each of the $O_2$ bond and antibond one-electron states (except for majority-spin $\pi$'s for which both are occupied) when treating the oxygen molecule interacting with a LSM surface. This includes the $\sigma$ bonds, which enter with the same formula, though they make negligible contribution to the polarizability. We should also note in passing that the same formula should apply to nitrogen molecules, with slightly smaller spacing than $d_O$, but both π bonds contributing.

5. An Oxygen Atom Over LSM

It will be helpful first to treat an individual atom, rather than a molecule, above a Mn ion in a SrMnO₃ substrate, with the oxygen levels shown to the far right Fig. 1. In the cluster levels the coefficient of the Mn $d$ orbital is $1/\sqrt{2}$, so the coupling with the oxygen orbitals is reduced to $V_{pdm}/\sqrt{2}$. For this calculation we need also to include the orthogonality shifts of these levels, and the repulsion between the oxygen atom and the substrate using Eq. (4). For each of the four cases (up- and down-spin, π and σ), we need to solve a three-by-three Hamiltonian (oxygen level, bond and antibonding cluster states), with the appropriate $U^*$ added to each state initially empty. The highest occupied state for each case are shown in Fig. 2, along with the total energy, which is the sum of the energies of occupied levels plus extra repulsions.

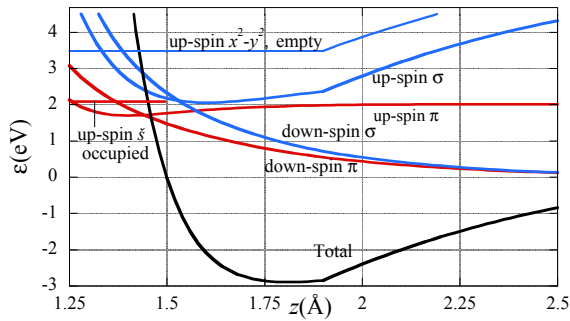

Fig. 2. Highest occupied states of each category for an O atom, as a function of the distance $z$ to an Mn ion under it in the substrate. The total, in black, includes additional repulsions, is a minimum near 1.8 Å, and is measured from the energy at large distances. The cusps at 1.90 Å arise from $U^*(Mn) = 0$ for distances less than that, and are not physical. The lowest empty state is also shown.

The result is very informative. The minimum total energy, −2.89 eV at $z = 1.8$ Å, indicates that the oxygen can be bound to the surface as a neutral atom. This may be the most important finding for the oxygen atom. The binding arose largely from the up-spin σ electron on the oxygen, which came down more than two electron volts in energy. We would call this a polar covalent bond; the oxygen atom has not acquired formal charge. Some might wish to think of it as an ionic bond, but that would seem to require naming oxygen the positive ion.

We did not find this deep minimum over a substrate oxygen ion. The repulsion remains but the dropping of the σ level to the substrate energy, which caused the minimum, is absent and the total energy increased monatonically with decreasing distance. Similarly, the behavior is quite different on LaMnO₃ (or a $Mn^{3+}$ site in $La_{1-x}Sr_xMnO_3$) in which the $3z^2–r^2$ cluster orbital, to which this σ state is coupled, is already occupied. It is more favorable for LaMnO₃ to reverse the spin of the incoming oxygen and seek analogous bonding with the π state. We did that calculation, with the occupied up-spin state in the atom, to the far right in Fig. 1, being a π state. The curves are quite different from Fig. 2, but the total energy has a similar minimum of –1.51 eV at $z = 1.86$ Å. Indeed the $Mn^{3+}$ site can bind a single oxygen, but not nearly as strongly as the $Mn^{4+}$.

An oxygen atom coming in over an $Mn^{4+}$ site with thermal kinetic energy, near zero, will be accelerated by the dropping total energy, acquiring a kinetic energy near 3 eV before being turned around near $z = 1.5$ Å and accelerated back outward, leaving the surface. With just what we have included so far, there can be no energy loss and no chance of capture of the atom. The fact that there was no barrier to reaching the position of a bound atom at 1.5 Å was not a sufficient condition for capture. We must look for energy-loss mechanisms.

## 6. Level Crossing

An interesting mechanism for energy loss by electron transfer can be seen in Fig. 2. The only empty level in the diagram is the up-spin nonbonding $x^2–y^2$ state at 3.5 eV + $U^*(Mn)$. We see that occupied σ states cross that level at small spacing, and we might ask if an electron transfer is possible. It would have to be the up-spin state or a spin flip would be required, and it occurs at such high total energy that it would not be expected here in any case. However, the possibility arises again later at lower energies in LSM and shows up in a number of our plots so we should discuss it. If a transfer did occur, with an electron left in this nonbonding state at high energy, that energy would be taken from the kinetic energy of the oxygen atom, now becoming a positively charged ion, which might then not have sufficient energy to leave. It could be bound to the surface, losing energy to lattice vibrations as we shall describe in the next section, presumably ending up at a site different from where the electron was left. There we would expect it to pick up an electron from the conducting substrate and remain bound to the surface as a neutral atom, at the –2.89 eV of Fig. 2.

This transfer of an electron at a level crossing is an intricate occurrence. For the ideal geometry we have assumed, the two levels have different symmetry, and no transfer can occur. However, if the incoming oxygen were displaced very slightly away from the axis of the cluster there would be a coupling between the two levels, and they would separate into an upper and a lower curve, with no crossing. Then a transfer would be guaranteed, at least for slowly moving levels. In fact when the coupling is small, both outcomes are possible. The probability that the system will transfer to the level it crosses, and is coupled to by $V_{1,2}$, is given in perturbation theory by[12]

$$P_{1,2} = \frac{2\pi V_{1,2}^{2}}{\hbar \partial(\varepsilon_1 - \varepsilon_2)/\partial t} \tag{10}$$

if the two levels change their energy relative to each other by $\partial(\varepsilon_1–\varepsilon_2)/\partial t$ at the time of the crossing. For our case this probability increases from zero only with the fourth power of the displacement of the trajectory from the symmetry axis, so there is a "sweet spot" around the symmetry axis where the electron would probably follow the curve of Fig. 2 across the nonbonding level as the oxygen arrived, and if it left outside of the sweet spot the electron would be transferred to the substrate on the way out. Similarly, if it missed the sweet spot on the way in, but left through it, the electron would again be left on the substrate. For the parameters we have used, the area of the sweet spots over the Mn ions is some 5% of the total area. Again, for the oxygen atom in Fig. 2 the crossing occurs at such high energy that we do not expect the question to arise for this case.

## 7. Phonon Generation

The generation of phonons of course is another possible mechanism for energy loss, with or without electron transfer. It can readily be estimated classically, and quantum effects are not expected to be important, using the total energy curve from Fig. 2, or a fit $E(z) \approx 5000/z^{10} - 44/z^4$ in eV if $z$ is in Å. To do this, we represented the crystal by a chain of ten atoms, alternately Mn and O, as shown to the right in Fig. 3. They are connected by springs, with constants $\kappa = 16$ eV/Å$^2$ fit to the bulk modulus, within the chain and to four lateral neighbors with the same $\kappa$. The result of such a classical dynamical calculation is shown in Fig. 3. Results were similar if we replaced the chain by a single atom, but then as energy is transferred into and out of the single mode (quantizing it might be numerically significant), the oxygen atom is soon kicked off; with ten atoms this takes much longer as the energy was distributed in many modes. At least for these arrival parameters we expect a sticking coefficient near one for a single oxygen atom. The result should be similar at other sites. If the oxygen atom had given up an electron as described above, a similar dissipation of the remaining energy would be expected to occur.

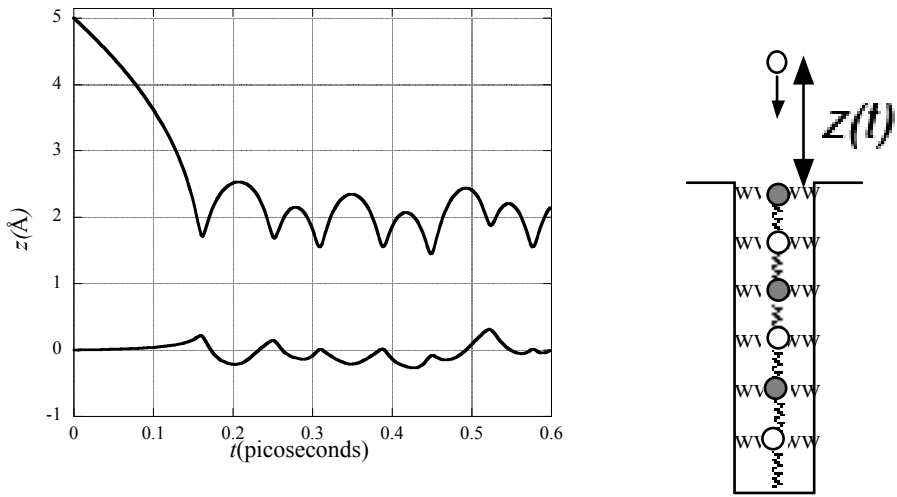

Fig. 3. The upper curve is the position of an O atom, initially approaching an Mn ion in a surface from directly above with 100 meV kinetic energy. The substrate was modeled as illustrated to the right, but with a chain of ten atoms, alternately Mn and O. Below is shown the position of the top (Mn) ion of the chain.

The large loss to vibrations arose only because of the considerable acceleration of the oxygen atom, to 3 eV, before striking the surface like a hammer. With only the repulsive term in our fit to the energy as a function of $z$, almost no vibration is excited as the oxygen atom bounces off the surface. In the case of an electron transfer, the loss to vibrations might be slow, but would eventually occur.

## 8. Oxygen Molecule on LSM, Perpendicular Orientation

The extension of this theory to oxygen molecules is quite straightforward, particularly with the molecular axis normal to the substrate surface. Cluster states shown in red in Fig. 1 are coupled only to molecular states shown in red, and the same for states shown as blue. For each category of level we now have a basis of *four* states rather than three, the upper and lower cluster levels and the bond and antibond on the molecule. The coupling between the cluster orbital and the levels on the nearest oxygen atom at a distance $z$ is reduced by another factor of the coefficient $1/\sqrt{2}$ of the nearest oxygen orbital in the bond and antibond states, and that affects the nonorthonality shifts through the same $\lambda = 1.44$. We also include the van-der-Waals interaction by adding $\delta U^*/4$ from Eq. (8) to each $O_2$ molecular level which enters the calculation. We first calculate the levels holding the oxygen spacing in the molecule at $d_O = 1.22$ Å, giving the highest-energy occupied levels and total energy shown in Fig. 4. The $U^*(Mn)$ calculation was based upon $r = z + d_O/2$ so the cusp occurs at $z = 1.29$ Å and does not show in the figure. Without the van-der-Waals interaction the total would have dropped monatonically to zero with increasing $z$, rather than showing the minimum of −0.015 eV at 3.9 Å. We also tried minimizing the energy with respect to $d_O$ at each $z$, but the total energy differences were small, −0.0003 eV and $d_O = 1.235$ Å at the minimum, where $z$ was 3.8 Å. For our problem it is adequate to keep $d_O = 1.22$ Å. We also checked the predicted charges on the O atoms, finding both very close to neutral for $z \geq 2$ Å so that they are not important.

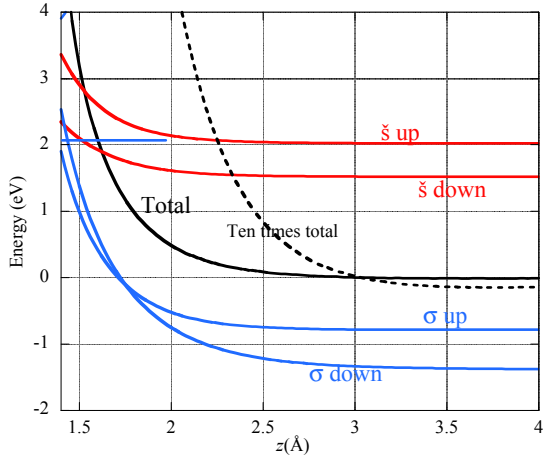

Fig. 4. The highest occupied levels for an $O_2$ molecule oriented perpendicular to the surface above a substrate $Mn^{4+}$ ion, the counterpart of Fig. 2 which was for an O atom. The total energy also includes the van-der-Waals interaction with the substrate, leading to the minimum −0.015 eV at 3.8 Å. The nonbonding level indicated at 2.02 eV is occupied; the empty $x^2-y^2$ nonbonding level is out of the figure at the top.

We note that any crossing of an empty level occurs to the left of the figure and is not accessible, as for the O atom. The bond at some 3.8 Å distance should be regarded as a van-der-Waals bond rather than a chemical bond. We repeated the dynamical calculation as in Fig. 3, fitting the total-energy curve by $230/z^8 - 1./z^3$ in eV if $z$ is in Å, and finding that there was not enough vibrational energy left behind to bind the molecule, even with an incident energy as low as 10 meV. The calculation held the oxygen spacing fixed, but with such small substrate effects we expect the molecular vibrations also to be negligible. A molecule above a $Mn^{3+}$ site (with La replacing some Sr ions) would have an additional down-spin antibonding σ orbital occupied, presumably reducing what little attraction was present, leaving again no significant phonon generation, nor sweet spots.

## 9. $O_2$ Oriented Parallel to a $SrMnO_3$ Surface

We redid the calculation with the molecular axis along an $x$ axis parallel to the surface, again centered over (at a distance $z$) a surface $Mn^{4+}$ ion of $SrMnO_3$. This was considerably more intricate, with for example the $zx$ cluster orbital no longer equivalent to the $yz$ cluster orbital. Also the states based upon the $e_g$ cluster states included both the $x^2-y^2$ and the $3z^2-r^2$ bonding and antibonding orbitals as well as the bonding σ and $z$-

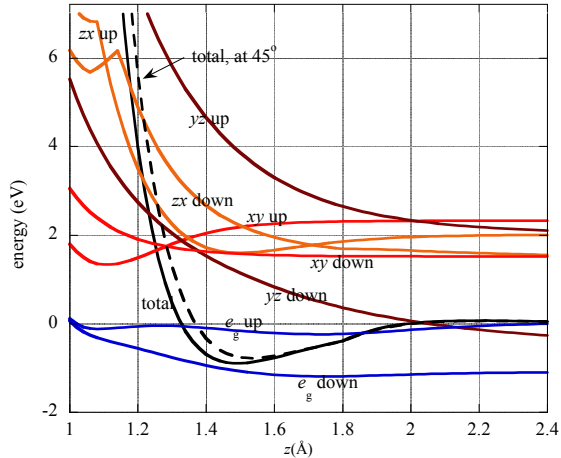

Fig. 5. The highest occupied levels for an $O_2$ molecule oriented parallel to the surface above a substrate $Mn^{4+}$ ion, the counterpart of Fig. 4 for perpendicular orientation. There are no nonbonding levels and the cusps for the $zx$ orbitals arise from a level crossing and are real. The total energy in black is for the $O_2$ axis along a cube direction in the surface; the dashed black line is if rotate 45° in the surface. The $xy$ and $e_g$ curves are also changed by the 45° rotation.

oriented π states of the molecule. We dropped the bonding cluster states for the $e_g$ case, which should have little effect on the energy, so that we again had only fourth-rank

Hamiltonian matrices to solve. The results, shown in Fig. 5, were a surprise. The most important point is that there is again a strong bonding interaction as for the *atom* incident on the surface. It leads to a deep minimum of −0.90 eV in the energy at $z$ = 1.49 Å. The van-der-Waals interaction was dominant; without it the total energy had a minimum of only −0.16 eV at 1.58 Å. This total is the sum of contributions to each category (all occupied levels, though only the top level of each type was shown in Fig. 5, and the repulsions with the substrate oxygen neighbors to the Mn). It was dominated by the $zx$ and the two sets of $e_g$ cluster levels. The major difference from the $O_2$ with perpendicular orientation is that the coupling of the $z$-oriented π levels of the molecule to these cluster orbitals (both the $zx$ and $e_g$ sets) includes the large $V_{pd\sigma}$. For the perpendicular orientation, that coupling entered only for the molecular σ levels which are very widely spit and contribute less. In addition, there is coupling of the cluster orbital to both oxygen atoms in the molecule, increasing the coupling by a factor of √2. With this deep minimum we found large depositions of vibrational energy, as expected, enough to capture the molecule, as in Fig. 3.

These calculations leading to Fig. 5 again held the molecular spacing at 1.22 Å, but we then sought the minimum energy with respect to that spacing. We found that indeed near the minimum in the total energy the optimum spacing was close to 1.22 Å, but at small $z$ the spacing grew, to 1.43 Å at $z$ = 1.2 Å. Also, at larger $z$ the spacing grew to 1.32 Å at $z$ = 2.5 Å, but then dropped back to the starting 1.22 Å.

Another surprise are the cusps appearing in the $zx$ curves in Fig. 5. They arise from level crossing[13], but occur again at too high a total energy (near 6 eV) to provide an accessible sweet spot. A large contribution to the high total energy came from the repulsion of the molecule by the oxygen substrate neighbors; we shall see in Fig. 7 that removing one oxygen neighbor reduces the energy by about 1 eV, still far too little to make the sweet spot accessible.

If we rotate the oxygen molecule by an angle θ in the plane parallel to the surface, there is some change in the bonding terms. For a π/4 rotation the $x^2$–$y^2$ and $xy$ orbitals are interchanged, so we need to make the corresponding changes in the program. There is another change from the modification of the repulsion with the substrate oxygen ions, which becomes minimum at orientation 45° from the cube axes. The van-der-Waals energy is not affected. We redid the total-energy calculation with these changes to obtain the dashed curve in Fig. 5. Actually this energy is 110 meV higher at π/4 and so the expected orientation of the oxygen molecule is parallel to a cube axis, bound to the surface of $SrMnO_3$. The increase in bonding energy was larger than the decrease in oxygen-oxygen repulsion.

We note finally, that if instead of bringing the molecule to the surface to be bound by −0.9 eV, we had split it into atoms, costing 5.2 eV, and brought the two atoms in to be bound as we saw in Section 5, gaining −2.89 eV each, the net energy would have been −0.58 eV. This is only 0.32 eV higher than the energy of the molecule, suggesting that dissociating the molecule on the surface takes only this small energy. These are small differences in large numbers so we cannot have confidence in the accuracy, but the picture is quite interesting. The molecules could be rather easily dissociated on the surface, with each atom bound to a different $Mn^{4+}$ site, but they could only boil off as molecules.

## 10. $O_2$ Oriented Parallel to a $LaMnO_3$ Surface

The only important difference if the $O_2$ is over a $LaMnO_3$ surface is that with a $Mn^{3+}$ ion there is an additional electron in an $e_g$ state. We can run the same program but with three, rather two, majority-spin $e_g$ states occupied. Then also the $U^*(Mn)$ is added only to the upper, $x^2-y^2$, basis state. The result, shown in Fig. 6, provided another surprise. As the molecule comes in, the overlap interaction of the up-spin $3z^2-r^2$ orbital with the molecule raises it above the $x^2-y^2$ state, causing a level crossing appearing as a cusp near $z = 1.9$ Å. [The cusp looks peculiar since it happens to occur just where $U^*(Mn)$ goes to zero (at $z = 1.8$ Å, or $r = 1.9$ Å) so the cluster level based upon the $x^2-y^2$ has a small artificial cusp of its own.] This produces a sweet spot, with a rearrangement of electrons primarily within the substrate. This extra electron in the antibonding state has also caused the total energy to rise to +0.63 eV so most molecules coming directly over an Mn would be expected to be reflected before reaching the sweet spot and not reach the very shallow minimum inside. On the other hand, a molecule displaced sufficiently from this sweet spot would not feel this repulsion because the coupling which caused the rise in the $3z^2-r^2$ state would be weak, and capture might occur. Even with this weak bonding, the molecule cannot dissociate easily at the surface as on $SrMnO_3$. If we first separated the molecule, costing 5.2 eV, and brought the atoms in gaining −1.51 eV each (Section 5), the resulting energy is much higher, at +2.18 eV.

If we rotated the molecule by $\pi/4$ in the plane we found that the level crossing did not occur and the energy rose monatonically, as the dashed line in Fig. 6. The behavior in this case is complicated, and needs to be explored in detail, best in the context of LSM. In the mixed crystal, $La_{1-x}Sr_xMnO_3$, a fraction $x$ of the surface sites have the deep bonding well shown in Fig. 5, and a path for molecules to approach the surface, with perhaps some ending in the shallow well over the $Mn^{3+}$ clusters, seen in Fig. 6. We shall see in the next section that this could be important.

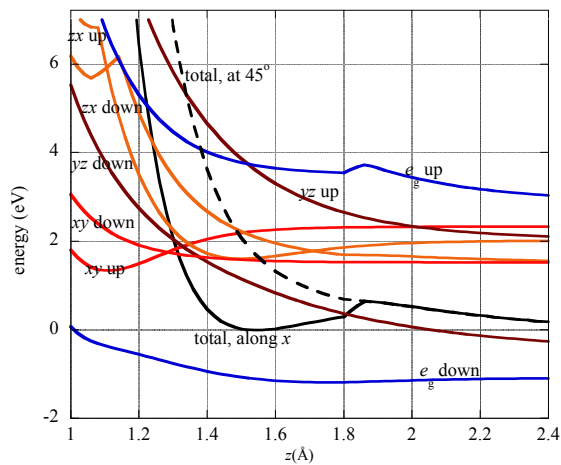

Fig. 6. $O_2$ molecule oriented parallel to the surface above a substrate $Mn^{3+}$ ion, the counterpart of Fig. 5 but with $Mn^{3+}$, rather than $Mn^{4+}$. There is a new crossing of up-spin $e_g$ levels near 1.85 Å. The minimum total energy is near zero at $z = 1.55$ Å. For a 45° orientation the level crossing does not occur and there is no minimum at all.

## 11. Surface Vacancies

The most important defect which may be present in the substrate is an oxygen vacancy in the surface plane. An oxygen vacancy in the bulk material has broken the bonds with its two Mn neighbors, leaving a formal plus-two charge. In $SrMnO_3$ we expect that charge to be balanced by converting two $Mn^{4+}$ ions to $Mn^{3+}$ ions, by placing an electron in a majority-spin $e_g$ antibonding state, just as when a $La^{3+}$ ion is substituted for a $Sr^{2+}$ ion. Similarly in $LaMnO_3$ the presence of a vacancy will cause two $Mn^{3+}$ ions to become $Mn^{2+}$ ions. These extra electrons will be attracted to the vacancy, with lowest energy with one in each of the two Mn sites adjacent to the vacancy, producing a *neutral* vacancy complex which can diffuse through the lattice. This is in contrast to the diffusion of the positively charged vacancy in the $ZrO_2$ electrolyte where substituting some Y for Zr produces a net charge of $-e$, which is compensated by creating an oxygen vacancy of charge $+2e$ for every two dopant Y's. In a fuel cell, this +2 vacancy diffuses through the $ZrO_2$ and then draws two electrons from the conducting LSM cathode as it enters. For this study, the important point is that the oxygen vacancies are essentially neutral species.

The state of the vacancy is little changed if it goes into a surface $MnO_2$ plane, where it again has two Mn neighbors in the surface plane. The energy gain by relaxation of neighboring ion positions may be larger for the free surface, which may favor surface segregation of the vacancies, but a quadrapolar field from the vacancy might favor interior sites.

Then, if vacancies are present in the surface plane we can imagine that one of the neutral molecules on one of the Mn neighbors, initially oriented along a [100] cube direction, might roll over, placing an oxygen atom into the neutral vacant site. This would form a standard $O^{2-}$ constituent of the crystal, leaving the second neutral atom bonded to the surface, as we described in Section 5. It would then be ready to fill another vacant site which diffused by. Similarly, if the oxygen molecule had dissociated at the surface, at the cost of 0.32 eV (Section 9), a neutral atom on a neighbor site might easily roll into the vacant site in the same way.

If this were $SrMnO_3$ we expect the two neighboring sites to be $Mn^{3+}$ with the extra electron, so the molecule would be described by the total energy of Fig. 6, not the strongly bound molecules described by Fig. 5. Similarly, an oxygen *atom* would be bound to the substrate by $-1.51$ eV, rather than the $-2.89$ eV for the $Mn^{4+}$ site (Section 5). In the case of $LaMnO_3$, if the vacancy has the $Mn^{2+}$ neighbors we anticipate a uniform repulsion of the molecule. However, having the vacancy neighbor eliminates one repulsion of about 1 eV for a molecule or atom directly over the Mn, and more if it shifts toward the vacancy. We must look more carefully at the site next to the vacancy and the rolling over of the $O_2$ molecule. We do that for a $Mn^{3+}$ site, and for a $Mn^{4+}$ site.

## 12. The Roll-Over

Consider a molecule over a Mn cluster, as illustrated to the right in Fig. 7. It is good to begin with full shells, though we have seen that a molecule is not bound to such a $Mn^{2+}$ site (with a full shell of majority-spin $d$ electrons and an empty shell of minority-spin $d$ electrons). With full shells we would expect the cluster to appear quite spherically symmetric to the molecule, except for the repulsion by the oxygen neighbors in the surface. For $Mn^{3+}$, where we found a weak binding of the molecule, there was a single empty up-spin state, of symmetry $x^2-y^2$, providing some binding and the principal asymmetry, We expect the effect of coupling between the molecule and this one empty level, along with the oxygen repulsion, to provide the principal variation of the energy with θ as the molecule rolls over. We simply subtract the contribution of that state to obtain the variation with θ we seek. To be sure, when the molecule moves off axis there is coupling between both the $x^2-y^2$, the $3z^2-r^2$, the bonding and antibonding σ levels of the oxygen and the $z$-oriented bonding and antibonding levels so a full calculation would be very intricate. However the principal coupling at small displacements of this up-spin $x^2-y^2$ antibonding level at 3.49 eV is with the empty antibonding σ and π antibonds of the molecule. These empty antibonding molecular states are shifted by $U^*(O)$ and if we keep just these three orbitals the calculation is the same as for the many levels we have treated, except for a somewhat intricate determination of the couplings in this low symmetry. This is a considerable simplification, but we shall find that the bonding term is small compared to other terms, and so this seems adequate. The results are shown in Fig. 7. For the full substrate the energy variation with angle of roll is dominated by the repulsion with the oxygen substrate ions in the plane of the roll. The effect of the bonding with the empty $x^2-y^2$ cluster state is quite negligible. The same is true for an $Mn^{4+}$ ion, where the empty $3z^2-r^2$ orbital also contributes, as shown by the dashed blue line. Removing the neighboring substrate oxygen to the right, to form a vacancy, leads to the dashed black curve. The molecule simply rolls over with no barrier to overcome. We expect it to fill the vacancy and for the upper oxygen to shift back to a position over the Mn.

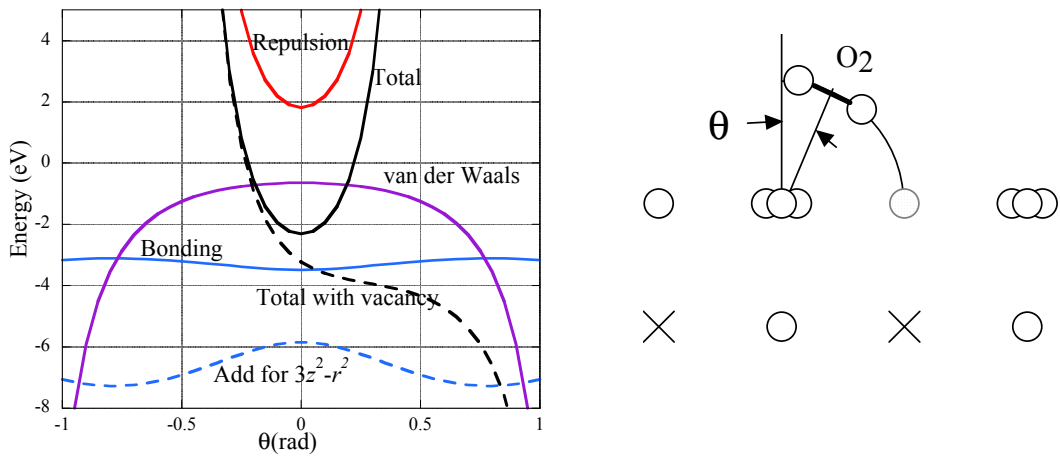

Fig. 7. Variation in energy as an $O_2$, initially parallel to the substrate rolls by an angle θ over a substrate $Mn^{3+}$ ion as shown to the right. The full curves show the contributions. The blue curve is the estimated contribution of the empty $x^2-y^2$ cluster orbital. For an $Mn^{4+}$ ion the contribution for the dashed blue curve would be added. The dashed black curve results if the repulsion to the substrate oxygen on the right is eliminated.

We may expect similar behavior for an oxygen atom on a site next to the vacancy, and the same approach should be appropriate. We have not yet carried out the calculation.

## 13. Discussion

In summary, the picture which emerges is much different than we anticipated. There are ample opportunities for capture of $O_2$ molecules, as well as O atoms, on planar surfaces of $La_{1-x}Sr_xMnO_3$. There may, however, be adsorbed $N_2$ molecules competing for the same sites. The neutral $O_2$ molecules are rather strongly bound, by about 0.9 eV, by van-der-Waals forces over $Mn^{4+}$ ions. The molecules are predicted to be oriented parallel to the surface, with axes along cube directions. There is also a very weakly bonded position over $Mn^{3+}$ ions, and a uniform repulsion over $Mn^{2+}$ ions. However, much of the weakness of that bonding came from repulsion with neighboring substrate oxygen ions, and a stronger bonding should be possible at an edge site or beside a vacancy. We found that these molecules could be dissociated on the surface for a net cost of only 0.32 eV, leaving the atoms strongly bound at $Mn^{4+}$ sites but more weakly at $Mn^{3+}$ sites.

We find that an oxygen vacancy from a zirconia electrolyte below the surface would pick up two electrons to become neutral when entering the LSM cathode, and could diffuse to the surface as a neutral vacancy complex. As such, its two neighboring Mn ions would be $Mn^{3+}$ in $SrMnO_3$ with some also $Mn^{2+}$ in the mixed LSM, so oxygen molecules and atoms tend not to be bonded to its neighbors. If however they were, perhaps because of the missing repulsion from the missing oxygen, they would roll into the vacant site, eliminating the vacancy, in the case of a molecule leaving the other oxygen bonded to a neighboring site.

These results seem consistent with what is known about related systems. Baniecki, et al.[2], find neutral water and $CO_2$ molecules bound with similar one eV energies to $SrTiO_3$ surfaces, both from Thermal Desorption Spectra (TDS) measurements and from accompanying density-functional calculations. They also find ( J. D. Baniecki, private communication) that if they generate surface vacancies, water molecules will leave an oxygen atom to fill the vacancy, releasing a neutral hydrogen molecule, similar to the roll-over process we envisage here for an $O_2$ molecule.

The closest we could find for such measurements related to TDS on LSM were by Kan, et al[15], which however focused on exchange of oxygen isotope tracers. It may be helpful to put numbers in our results to see more clearly what we would predict for such TDS measurements on the manganites. For the oxygen gas we might take the kinetic energy associated with each direction of motion as ½ $kT$. Then we may readily estimate the rate molecules hit a unit cell $(2d)^2$ in area as $r_H = 1.68 \times 10^7$ $P/\sqrt{T}$ per second if $P$ is in Torr and $T$ is in °K. This is also the rate a $Mn^{4+}$ cluster would acquire a molecule if it did not have one and if the sticking coefficient were one. We could also estimate the rate a molecule would escape from being bound by 0.9 eV to a site as $r_E = 1.59 \times 10^{13}$

exp($-10450/T$) per second if $T$ is in °K. [We took the attempt frequency for the leading factor from our estimate of an oxygen vibration frequency in LSM from Ref. 9.] Then solving the transition-rate equation,

$$\partial f/\partial t = r_E f - r_H(1-f),  \qquad (11)$$

for the steady-state occupation of a site we obtain an occupation of $f = r_H/(r_H + r_E)$, plotted in Fig. 8. The thermal desorption, for very slow ramping of temperature, would simply be the derivative of such a curve with respect to temperature, shown as the dashed line in the figure. The peak occurs near the temperature where $f = ½$, estimated by equating the formulae for $r_H$ and $r_E$ and solving numerically for $T$. Redhead[14] has given formulae for different ramping rates, but it would seem as well to simply integrate Eq. (11) over time for the $T(t)$ of the experiment and plot $\partial f/\partial t$ as the predicted TDS curve. In Fig. 5 of Ref. 15 is a temperature programmed desorption curve which shows an increasing desorption in 100 Torr oxygen at 800°K which the authors associate with the beginning of a desorption peak. That could be related to the peak we would obtain for 100 Torr from Fig. 8 at slightly lower temperature, but it is not convincing support. More complete data would give a direct check on our estimate of the 0.90 eV for the binding of the $O_2$. It would also tell if there were $N_2$ molecules adsorbed, as we expect. Any oxygen atoms bound to the surface would require too much energy to escape, except by first forming a molecule.

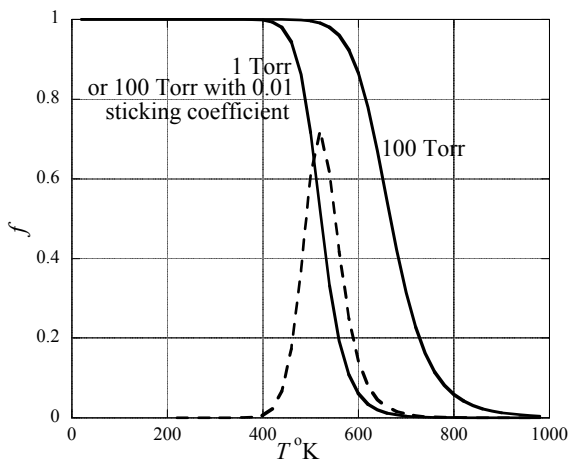

Fig. 8. The steady-state occupation of $Mn^{4+}$ sites by $O_2$ molecules, as a function of temperature for pressures of 1 Torr and 100 Torr, for unit sticking coefficient. The dashed curve shows the predicted thermal desorption spectrum (arbitrary units, proportional to $\partial f/\partial T$) for 1 Torr.

   Though the available experimental information to directly test our findings is very limited, it *is* known that the incorporation of oxygen is slow, here interpretable as the result of the molecules avoiding the sites neighboring any oxygen vacancy, or the surface being saturated with $N_2$.

Also, Fister, et al.[16] using total-reflection x-ray fluorescence have found segregation of Sr near the surfaces of LSM. Qualitatively we might expect this because of the strong binding of oxygen at $Mn^{4+}$ sites, the number of which increases with increasing concentration of Sr. That does not however fit with other observed trends. With the surface-oxygen mechanism, we would expect the segregation to be proportional to the *f* of Fig. 8, generally increasing with partial pressure, while they find decreasing segregation with increasing pressure. They also find an enthalpy of segregation very small on the scale of our 0.9 eV, which decreases with increasing oxygen pressure. This suggests another mechanism, and Fister, et al.[16], have suggested that oxygen vacancies are involved. Indeed two Sr ions could move into the region of each vacancy in $LaMnO_3$, to avoid converting two $Mn^{3+}$ ions to $Mn^{2+}$ions. This attraction would decrease with increasing oxygen pressure which reduces the number of vacancies, consistent with the direction of their trend. The validity of the picture presented here needs further experimental test, and TDS would seem to best choice, checking for the strongly bound molecular oxygen and nitrogen on the surface.

We should discuss finally the suggestion made in Ref. 6 that the slow incorporation of oxygen in metals, and other systems, arises from selection rules associated with the triplet state of the oxygen molecule. The motivation for the suggestion, based upon molecular reactions, can be understood using the same assignment of majority and minority spins we have used in this study. An oxygen molecule, with both antibonding $\pi$ states occupied only by electrons of spin down, might react with two CO molecules, both with equal numbers of up- and down-spin electrons. However the final products, two $CO_2$ molecules have equal numbers of up- and down-spin electrons so one electron needs to be flipped in the reaction, requiring spin-orbit coupling and greatly slowing the reaction. This is not inconsistent with any of our analysis because our final states were in all cases of the same total spin as the initial states. If a molecule rolled into the vacancy, as in the preceding section, the remaining adatom would retain the triplet spin of the starting molecule if the two spins on the Mn neighbors to the vacancy had been antiparallel, and no spin-flip is required.